\documentstyle[12pt]{article}
\def\be{\begin{equation}}
\def\ee{\end{equation}}
\def\ba{\begin{eqnarray}}
\def\ea{\end{eqnarray}}
\def\ga{\mathrel{\raise.3ex\hbox{$>$\kern-.75em\lower1ex\hbox{$\sim$}}}}
\def\la{\mathrel{\raise.3ex\hbox{$<$\kern-.75em\lower1ex\hbox{$\sim$}}}}
\newcommand{\sect}[1]{\section{#1}\setcounter{equation}{0}}

\newcommand{\fr}[2]{\frac{#1}{#2}}

\begin{document}

\begin{titlepage} 
\rightline{CERN-TH/2002-071}
\rightline{UMN--TH--2048/02}
\rightline{TPI--MINN--02/08}
\rightline{SUSX-TH/02-009}
\rightline{hep-ph/yymmnn}
\rightline{April 2002}  
\begin{center}

\vspace{0.5cm}

\large {\bf On the Stabilization of the Size of Extra Dimensions}\\[3mm]
  
\vspace*{0.5cm}
\normalsize

{\bf  Panagiota Kanti$^{\,(a)}$, Keith A. Olive$^{\,(b)}$} and 
{\bf Maxim Pospelov$^{\,(c)}$}

\smallskip 
\medskip 

$^{\,(a)}${\it Theory Division, CERN, CH-1211 Geneva 23, Switzerland}
 
$^{\,(b)}${\it Theoretical Physics Institute, School of Physics and
Astronomy,\\  University of Minnesota, Minneapolis, MN 55455, USA} 

$^{\,(c)}${\it Centre for Theoretical Physics,CPES,
University of Sussex, Brighton \\BN1~9QJ, United Kingdom }
\smallskip 
\end{center} 
\vskip0.6in 
 
\centerline{\large\bf Abstract}

We derive a general formalism for constructing a four-dimensional 
effective action for the radion field in $5$ dimensions, taking into
account the possible dependence of the scale factor on the internal
coordinate. First and second variations of the action  with respect to the
size of the extra dimension reveals two types  of algebraic constraints
on the combination of brane and bulk Lagrangians  that extremize the
action, and provide an effective mass for the radion field. We derive a
compact formula for the radion mass squared in terms of brane and  bulk
potentials for the case of a bulk scalar field. We further specialize 
these constraints for several, five-dimensional realizations of 
brane world models.

\vspace*{2mm} 

\end{titlepage}

\sect{Introduction}

Theoretical models with internal compactified space-like dimensions 
play a special role in string theory \cite{HW}, particle physics 
\cite{reviewpp}, and cosmology \cite{ADD,RS,cosmo1}. To be consistent 
with phenomenological requirements, the volumes of internal dimensions 
have to be stabilized, i.e. the corresponding moduli, or radions, 
have to acquire a regular mass term. Depending on the model at 
hand, a mass for a radion field may vary in a wild range from 
$10^{-3}$ eV to the Planck scale. The stabilization of extra dimensions 
is a phenomenological necessity both in models with a smooth distribution 
of matter in the extra dimensions and in models with branes, where the
metric  may acquire a strong dependence on the coordinates of the extra
dimensions,
$a = a(y,z,..)$. One of the most interesting recent developments for
model-building is the idea that the 
strong dependence of the scale factor on the coordinates in extra dimensions 
may have profound consequences for the gauge hierarchy problem
in particle physics \cite{RS,hierarchy}. 

The presence of matter in these models, either placed on 
the brane or smoothly distributed in the bulk, usually has a linear 
coupling to 
the radion field, displacing the radion from its vacuum value, and
creating the 
$T^n_n$ pressure-like components of the energy-momen\-tum tensor, 
where $n$ labels the
internal dimensions. The presence of this extra-dimensional pressure 
enables one to recover the usual $4D$ cosmological expansion \cite{kkop0}
and cures the crisis in  ``brane cosmology'' \cite{cosmo1} that consists
of a drastically different  cosmological evolution at late times. The firm
connection between the  stabilization 
mechanism and extra-dimensional pressure was further exemplified in 
Refs. \cite{csaki,kkop}. 
 
The static solution for the radion field equations ($\dot b =0$) implies
that the energy density, ``normal'' $4D$ pressure of matter and the 
extra-dimensional pressure $T^n_n$ must be related by an integral condition.
The first of a series of conditions was found in Ref. \cite{E}, in the
context of  the Randall-Sundrum model, and presented as a topological
constraint necessary for the consistency of the particular solution.
A similar integral condition, suitable for a non-warped case, $a(y) = const$,
was found in Ref. \cite{kkop} and it was shown to be associated with
the stabilization of the size of extra dimension. Finally, a whole family of 
different sum rules was presented in Ref. \cite{GKL}, applicable to the case 
of matter that is uniformly distributed ($x^\mu$-independent) in ``our''
dimensions. One of these sum rules was shown \cite{Papazoglou} to provide 
a consistency check for solutions with a constant radion field ($\dot b =0$).
The same sum rules were generalized in 6 dimensions in \cite{kmo}
and in d dimensions in \cite{Leblond}.

This paper extends the analysis given in  Ref. \cite{kkop} and includes the
case where the scale factor depend on the extra-dimensional
coordinate of a 5D spacetime. This results in a constraint on the
components of the  higher-dimensional energy momentum tensor, the {\it
extremization constraint}, that generalizes the simplest constraint
obtained in
\cite{kkop}.  Using the fact that the variation over $b$ with branes
at fixed positions,
$y_1$ and $y_2$, is equivalent in fact to the variation over the length
of  the extra-dimensional circle, we reduce integral constraints to a
set of  algebraic relations between brane Lagrangians and the bulk
Lagrangian,  calculated at the position of the branes. 

The presence of a stabilization mechanism leads to the radion 
relaxation to its minimum at $b=b_0$, with $\dot b = 0 $. 
However, the inverse is not true. The existence of a solution with 
 $\dot b = 0 $  might be a result of a fine-tuning that does not 
lead to a physically viable ``brane model''. Indeed, small 
perturbations of the solution around the extremal point $b=b_0$ may 
prove to be unstable. Such an instability can manifest itself as a 
tachyonic mass and/or a ghost-like signature for the kinetic term of 
the radion. To decide whether a particular brane world scenario is stable 
against radial perturbations or not, one has to find a close family 
of solutions around $b=b_0$ and expand in $b-b_0$. In many existing 
brane-world scenarios such an analysis is possible, whereas in some models 
that specify $a(y)$ and $T^m_n$ only  at a given point 
$b=b_0$, the stability cannot be analyzed. In cases when it is possible, 
the analysis of the stability ($\ddot b /b <0$) can be done with the use of 
Einstein's equations \cite{einstein} or directly in the action
\cite{GW,action}.
We choose the latter method, and obtain a rather compact algebraic 
formula for the radion mass squared, again in terms of 
brane parameters and bulk quantities calculated at the position of the branes. 
The {\it stabilization constraint} can be regarded as the condition 
on the positivity of the radion mass squared.

After obtaining the generic forms of the $\dot b =0$ and $\ddot b /b <0$
constraints, we apply them for some particular realizations of
brane-world models. We exemplify how the {\it extremization constraint}
($\dot b =0$) is satisfied in a number of static, five-dimensional
models, namely, in the Randall-Sundrum model \cite{RS}, in the two-brane
model  with a Casimir force \cite{mukho,hkp}, and in the two-brane model
with a classical massless bulk scalar \cite{kop1}. In the above cases, 
we also analyze the {\it stabilization constraint} and determine whether, or
under which assumptions, the radion has a positive or negative mass
squared. The results of this paper can be used to check the consistency of 
various brane model solutions and present a useful formalism that helps 
to analyze the stability of a particular extra-dimensional
solution against small perturbations of the radius of the internal spacetime. 

This paper is organized as follows. Section 2 presents the complete 
study of the $5D$ case addressing the generic form of the {\it stabilization}
and {\it extremization constraints}. Section 3 applies the derived constraints
to particular brane-world models. Section 4 presents our conclusions.

\bigskip

\sect{The radion potential in 5-dimensions}

Here, we assume that the brane self-energies, $V_i$, do not depend on the
four-dimensional coordinates either explicitly or implicitly. Therefore,
$V_i$'s can be, at most, functions of the extra coordinate $y$.
In this case, we may write down the following ansatz for the
five-dimensional spacetime
\be
ds^2=a(t,y)^2 (-dt^2 + \delta_{ij}\,dx^i dx^j) + b(t)^2 dy^2\,
\label{metric}
\ee
where $a(t,y)$ stands for the conformal factor multiplying a 
four-dimensional, flat line element and $b(t)$ denotes the
scale factor of the extra dimension.

We first focus on the gravitational part of the theory. For the above
ansatz, the 5D scalar curvature takes the form
\be
\hat R = R^{(4)}(t,y) -\frac{2}{b}D_\mu D^\mu b + \hat R^{(y)} = 
\frac{6 \ddot{a}}{a^3} -\frac{2}{b}D_\mu D^\mu b
+ \biggl(-\frac{12 a'^2}{a^2 b^2} -\frac{8 a''}{a b^2}\biggr)\,.
\label{scalar1}
\ee
The last two terms inside brackets, denoted as $\hat R^{(y)}$, are the 
part of the five-dimensional scalar curvature involving solely derivatives
with respect to the extra coordinate $y$. The covariant derivative of the
scale factor $b(t)$ will turn out to be a total derivative in 4D spacetime,
and hence drops out. The scalar curvature $R^{(4)}(t,y)$ is still
$y$-dependent and, therefore, the purely time-dependent, four-dimensional
scalar curvature ${\cal R}^{(4)}$ needs to be extracted. For this reason,
we define $a(t,y=0) \equiv a_0(t)$ and write the conformal factor as
\be
a(t,y)=a_0(t)\, \tilde a(t,y)\,
\label{def}
\ee
where $\tilde a(t,y=0)\equiv 1$, by definition\footnote{Note
that, in the case of static branes, this definition is not necessary as $a_0(t)$
reduces to a constant.}. $a_0(t)$ will play the role of
the 4D metric component that we need for the effective theory and so we write
\be 
R^{(4)}(t,y)= \frac{1}{\tilde a^2}\,{\cal R}^{(4)} - \frac{6}{\tilde a^3
a_0^4}\,
\partial_\mu[a_0^4 \,\partial^\mu \tilde a]\,,
\label{scalar2}
\ee
with ${\cal R}^{(4)}=6 \ddot a_0/a_0^3$.

We now turn to the action of the five-dimensional gravitational theory, which
can be written as
\be
S=- \int d^4x\oint dy\,\sqrt{-G}\,\biggl\{-\frac{\hat R}{2 \kappa^2_5}
+ {\cal L} \biggr\}\,
\label{action1}
\ee
where $\kappa^2_5$ stands for the five-dimensional Newton's
constant and ${\cal L}$ is a generic 
Lagrangian that may include bulk cosmological constants, brane-self energies and 
various bulk and brane matter fields. The integral over $y$ is performed over 
a compact dimension. We can recover finite volume, but infinite extra
dimension type models by sending one of the branes off to infinity. In
case when two infinitely thin branes are located at 
$y_1$ and $y_2$, the domain of integration over $y$ can be decomposed into three
separate pieces that represent the integral over the bulk and the
integrals across  the branes:
\be
\oint = \int_{y_1}^{y_2} +  \int_{y_2}^{y_1} +
\int_{y_1-\epsilon}^{y_1+\epsilon} + 
\int_{y_2-\epsilon}^{y_2+\epsilon}\,.
\label{domain}
\ee

The part of the above action consisting of the five-dimensional scalar curvature,
$\hat R$, will provide the kinetic part of the four-dimensional effective
action describing the dynamics of the graviton and the radion field.
By using the expressions (\ref{scalar1}) and (\ref{scalar2}) and integrating
by parts, the gravitational part takes the form
\ba
S_G &=& \int d^4x\oint dy\,\sqrt{-G}\,\frac{\hat R}{2 \kappa^2_5}\nonumber \\[1mm]
&=& \int d^4x\,\sqrt{-g_4}\,\Biggl\{\,\biggl(\oint dy\,\tilde a^2 b\biggr)\,
\frac{{\cal R}^{(4)}}{2\kappa^2_5} + \frac{3}{\kappa^2_5}\,\oint\,dy
\,\partial_\mu(\tilde a b)\,\partial^\mu \tilde a +
\oint dy\,\tilde a^4 b\,\frac{\hat R^{(y)}}{2\kappa^2_5}\,\Biggr\}\,,
\ea
with $\sqrt{-G}=a_0^4\,\tilde a^4 b$ and $\sqrt{-g_4}=a_0^4$. 
In order to obtain standard 4D Einstein gravity 
we perform a conformal transformation of the
four-dimensional metric tensor in order to eliminate the coupling between
the radion field and the scalar curvature ${\cal R}^{(4)}$. We set:
\be
(g_4)_{\mu\nu}=\frac{1}{A(b)}\,(\bar g_4)_{\mu\nu} \,\,\Rightarrow
\,\, \sqrt{-g_4}=\frac{1}{A^2(b)}\,\sqrt{-\bar g_4}\,,
\label{conformal}
\ee
where
\be
A(b)=\frac{\kappa^2_4\,b}{\kappa^2_5}\,\oint dy\,\tilde a^2(t,y)\,, 
\ee
which leads to the result
\be
{\cal R}^{(4)}=A(b)\,\biggl[\bar {\cal R}^{(4)} + 3 D_\mu D^\mu \ln A(b) -
\frac{3}{2}\,\partial_\mu \ln A\,\partial^\mu \ln A \biggr]\,.
\ee

Then, the entire action can be written as,
\be
S= -\int d^4x\,\sqrt{-\bar g_4}\,\biggl\{-\frac{\bar {\cal
R}^{(4)}}{2\kappa^2_4} + B(b)\,\partial_\mu b\,\partial^\mu b + \bar
V_{eff}(b)\biggr\}\,,
\ee
with
\be
B(b)= -\frac{3}{\kappa^2_5}\,\frac{1}{A(b)}\,\frac{\Bigl(\oint\, dy\,
\partial_\mu(\tilde a b)\,\partial^\mu \tilde a\Bigr)}
{\partial_\mu b\,\partial^\mu b}
-\frac{3}{4\kappa^2_4}\,\Bigl(\frac{\partial \ln A}{\partial b}\Bigr)^2\,,
\ee
and
\be
\bar V_{eff}(b)=\frac{1}{A^2(b)}\,\oint dy\,\tilde a^4 b\,
\Biggl\{- \frac{\hat R^{(y)}}{2 \kappa^2_5}  
+ {\cal L} \Biggr\}\,.
\label{eff1}
\ee
Note that, having restored the four-dimensional gravitational term,
all remaining terms in the action will constitute the building blocks of the
effective action for the radion field. 
Therefore, terms involving $x^\mu$-derivatives of the auxiliary functions
$\tilde a$ and $A(b)$ are bound to contribute to the kinetic term of the
only remaining  degree of freedom in the theory, that is of the radion field,
while any other terms will contribute to its effective potential.

At this point, we are able to obtain the constraints on the combination of
the scale factors and components of the stress-energy that were previously
derived in \cite{kkop} and \cite{GKL}. These constraints were derived under
the assumption that $\delta \tilde a / \delta b = 0$, that is, that the
perturbation with respect to the value of the radion field does not affect
the value of the scale factor. In this case, $A(b)$ is simply proportional
to
$b$ and the variation of the action (\ref{action1}) takes the following
simple form:
\begin{eqnarray}
\delta S &=& - \oint d^4 x\,dy \left[ -\frac{\delta (\sqrt{-G}\,\hat R)}
{\delta G_{MN}}\,\frac{\delta G_{MN}}{2 \kappa^2_5} -
\frac{\sqrt{-G}}{2}\,T^{MN}\,\delta G_{MN} \right] \nonumber \\[1mm]
&=&
- \oint d^4 x\,dy \left[ \delta \left(\frac{\sqrt{-\bar g_4}\,\tilde a^4}
{2\kappa_5^2\,b A^2}\right)
\left[\fr{8 a''}{a} + \fr{12 a'^2}{a^2}\right ]
-\fr{1}{2}T^5_5a^4 \fr{\delta b^2}{b} - \fr{1}{2} T^\mu_\mu a^4b
\fr{\delta(a^2)}{a^2} \right ]\\[1mm] \nonumber
&=& - \oint d^4x\,dy\,\sqrt{-G}\left[ -\fr{12 a''}{\kappa_5^2b^2a} + 
T^\mu_\mu - 2 T^5_5 \right]
\fr{\delta b}{2b}\,.
\label{kkop-gkl}
\end{eqnarray}
Note that the variation in the quantity $b^2 {\hat R}$ vanishes with
these assumptions. In the above expression we performed an integration by
parts and used the standard definition for the stress energy tensor. It
is easy to see that the square brackets  in the second line of Eq.
(\ref{kkop-gkl}) simply contains a combination  of the $ii$ and $55$
components of Einstein's equations, 
$ R^M_N -\fr{1}{2} G^M_N R^K_K - \kappa_5^2 T^M_N = 0 $. Obviously, 
Eq. (\ref{kkop-gkl}) must vanish identically when $a(y)$ is the solution. 
Integrating by parts the term with $a''$ we obtain the following 
constraint:
\be
\oint d^4x\,dy\,\sqrt{-G}\left[ 
\fr{36 a'^2}{\kappa_5^2a^2} + T^\mu_\mu - 2 T^5_5 \right] = 0. 
\label{simple}
\ee
This constraint generalizes a formula from \cite{kkop} and includes the
dependence of
$a$ on $y$ (i.e. warping). The constraint  derived in \cite{kkop} is
valid only in the case of small $ T^\mu_\mu $.  Indeed, a non-trivial
dependence on $y$-coordinate usually arises  due to the presence of
branes. In this case, $a'/a$ is proportional to the  brane energy density
$\rho$. In this case, the first term in (\ref{simple}) is  just an
$O(\rho^2)$ correction and can be dropped, leading exactly to the
expression found in
\cite{kkop}. For 5-dimensional spacetimes with significant warping, the
constraint (\ref{simple}) is valid instead. A similar, but not identical,
constraint was derived in Ref. \cite{GKL} for solutions with large
warping by rearranging the components of Einstein's equations. The same
combination $T^\mu_\mu - 2 T^5_5$ appears in both constraints, however
Eq. (\ref{simple}) has an extra term, proportional to $a'^2/a^2$, and a
different coefficient in front coming from 
$\sqrt{-G}$. The apparent differences are due to the fact that
the two constraints were derived with different methods, however,
they are both valid under the same assumptions and, are therefore,
equivalent.

In the above derivation we assumed that $\tilde a $ and $b$ are 
totally independent, and this is why the form of the constraint comes 
out to be identical with the combination suggested by Einstein's
equations.  However, in order to study the stability of the physical
size of  the extra dimension, one can 
no longer assume that $\tilde a $ and $b$ are independent and that $A(b)$
is simply linear in $b$. Instead, we would have to consider $\tilde a$ as a 
function of $by$, $\tilde a = \tilde a (by)$ and the variation of action over 
$b$ might be a more complicated procedure. It is important to 
note that there is always a residual symmetry under a simultaneous rescaling of 
$y$ and $b$. All physical parameters must remain invariant under
$y \rightarrow Cy$ and $b \rightarrow b/C$, where $ C $ is an arbitrary 
constant. Below, we will consider a two-brane model 
with branes positioned at $y_1$ and $y_2$. Then, the extremal size of the 
extra dimension is given by certain fixed values of $y_1b_0$ and $y_2b_0$, 
which are invariant under the above symmetry. The variation over $b$ around
$b_0$ with fixed $y_1$ and $y_2$ is equivalent to a variation over the
length of  the fifth dimension. Moreover, using this rescaling
symmetry,  we can always choose $b_0=1$, i.e. absorb $b_0$ into the
definition of 
$y_1$ and $y_2$.

A canonically normalized radion field $r(t)$ is obtained by demanding that
the kinetic term $2B(b)\,\partial_\mu b\,\partial^\mu b$ can be written as
$\partial_\mu r\,\partial^\mu r$. For small deviations of the scale factor
$b(t)$ from a stable minimum, $b_0$, of its potential, the normalized
radion field may be written as:  $r(t)=\sqrt{2B(b_0)}\,[b(t)-b_0]$, 
if the radion is not a ghost-like degree of freedom, i.e. $B>0$.
Note that, in the pursuit of deriving {\it extremization} and {\it
stabilization constraints}, we can continue working with the metric
function $b(t)$ instead of the radion field $r(t)$ since: 
\be 
\frac{\partial \bar V_{eff}}{\partial r}\biggr|_{r_0}=
\Bigl(\frac{\partial b}{\partial r}
\Bigl)\,\frac{\partial \bar V_{eff}}{\partial b}\biggr|_{b_0}\,,
\qquad
\frac{\partial^2 \bar V_{eff}}{\partial r^2}\biggr|_{r_0}=
\Bigl(\frac{\partial b}{\partial r}\Bigl)^2\,
\frac{\partial^2 \bar V_{eff}}{\partial b^2}\biggr|_{b_0}\,.
\ee
Therefore, for a well defined relation between $r(t)$ and $b(t)$, the
transition from one field to the other makes no difference in the vanishing
of the first derivative of the effective potential. In the same way, 
stability of the extra dimension in terms of $b(t)$, that is 
$\partial^2_b \bar V_{eff}>0$, leads to stability in terms of $r(t)$,
$\partial^2_r \bar V_{eff}>0$, and vice versa. We note that in particular models, 
the value of $B(b_0)$ may also be very important because it leads 
to a rescaling of the mass term, and a very large value of $B(b_0)$
may lead to a phenomenologically unacceptably low value of the radion
mass. 

Before we go into explicit calculation of the variation  of the
action over $b$ with $b$-dependent $\tilde a$, we would like to 
specify some components of the matter Lagrangian that we consider here. 
We allow ${\cal L}$ to include the bulk Lagrangian and brane
self-energies that may depend on the value of bulk fields:
\be
\int dy \,\tilde a^4 b \,{\cal L} =
\int_{y_1}^{y_2} dy \,\tilde a^4 b \,{\cal L}_B + 
\int_{y_2}^{y_1} dy \,\tilde a^4 b \,{\cal L}_B +
\sum_{i=1,2} \tilde a^4(by_i) V_i + U(by_1,by_2)\,.
\label{Lcompose}
\ee
In this expression, the bulk Lagrangian may include a variety of fields, 
however, for the sake of simplicity we will restrict to the 
case of a single scalar field $\phi$
\be
{\cal L}_B = \fr{1}{2b^2}\left(\fr{d\phi}{dy}\right )^2 +V_B(\phi)\,.
\label{bulkL}
\ee
Note that the bulk cosmological constant $\Lambda_B$ has been absorbed into
the definition of $V_B$. We also allow the brane self-energies to depend
on the value of the same field $\phi$. A constant part of $V_i$ at the
extremal value $\phi=\phi(b_0y_i)$ is a usual brane self-energy, 
$\Lambda_i = V_i(b_0y_i)$. Like $\tilde a$, $\phi$ is always a
function of $by$, ${\cal L}_B= {\cal L}_B(by)$ and $V_i = V_i(by_i)$.

Finally, $U(by_1,by_2)$ in (\ref{Lcompose}) represents an effective interaction 
that may arise due to quantum effects in the final volume
and which often cannot be formalized in the language of the Lagrange density in 
$y$-space. Sometimes this piece of the potential by itself may lead to the
stabilization  of the extra dimension. For the remainder of the
calculations we would like to keep a generic form for this potential,
specifying its form only in particular applications. 

We now turn to the explicit calculation of the variation of action with 
respect to $b$. In order to do that, we perform a change of variables,
$\xi = b y$.  This change brings about a significant simplification, as
$b$  now  only  enters in the calculation through the limits of
integration, i.e. at specific points $y_1$ and $y_2$
\be 
A^2(by_1,by_2)\bar V_{eff} = 2 \int_{by_1}^{by_2} \tilde a^4(\xi)
\left [ - \fr{1}{2 \kappa_5^2}\fr{12 \tilde a'^2(\xi)}{\tilde a^2(\xi)} + 
{\cal L}_B(\xi) \right ] d\xi +
\sum_{i=1,2} \tilde a^4(by_i) V_i
+ U(by_1,by_2)\,.
\label{Vofxi}
\ee
where we have now assumed equally spaced branes on the circle.
In the above expression, and henceforth, primes denote differentiation
with respect to $\xi$. Note, 
that we used the integration by part to remove $\tilde a''$ from the 
$\hat R^y$-part of the effective potential. This allows us to remove the 
so-called Gibbons-Hawking boundary terms, i.e. singular terms in $\hat R^{(y)}$
when $a''$ is taken at the positions of the branes. When $b$ is taken at 
its extremum, $\bar V_{eff}$ can be interpreted as the effective
four-dimensional  cosmological constant. 

The first derivative over $b$ is computed trivially, and the result takes the 
following form:
\ba
\label{first}
&&A^2(by_1,by_2)\fr{d\bar V_{eff}}{db} = \\
\nonumber
&&-\left. \fr{4 \kappa^2_4}{\kappa_5^2}\,A y_i\tilde a^2(by_i)\bar V_{eff}
\right|_{i=1}^{i=2}
 + y_i\tilde a^4_i\left.\left [-\fr{12 \tilde a'^2_i}{\kappa_5^2
\tilde a^2_i} +2{\cal L}_B(by_i) \right]\right|_{i=1}^{i=2}+
\sum_{i=1,2} y_i \fr{\partial}{\partial(by_i)}\left[ V_i \tilde a^4_i + U\right],
\ea
where $a_i \equiv a(by_i)$. The terms with $V_i$'s can be further
expanded as
\be
 \fr{\partial}{\partial(by_i)} V_i \tilde a^4_i = 
\tilde a^4_i \left[ 4 \fr{\tilde a'_i}{\tilde a_i}\,V_i + 
\fr{dV_i}{d\phi} \phi'_i\right ].
\ee
In this expression, the signs of $\tilde a'_i$ and 
$\phi'_i$ are defined according 
to the following rule:
\ba
\label{defini}
&~& \tilde a'_1 = \fr{d \tilde a}{d\xi}\biggl|_{(by_1 + 0)}\,, \\[2mm]
&~& \tilde a'_2 = \fr{d \tilde a}{d\xi}\biggl|_{(by_2 - 0)},
\ea
with similar definitions for $\phi'$. In other words, the derivatives are taken 
on the right side of brane 1, and on the left side of brane 2 since we assumed
$y_1<y_2$.

To derive the {\it extremization constraint}, we put $b=b_0=1$, and 
use equations of motion for the scale factor and $\phi$, as well 
as junction conditions on the branes, both for the scale factor and 
the scalar field:
\ba
\label{boundary}
\fr{\tilde a'}{\tilde a}(y_1) = - \fr{\kappa_5^2}{6} V_1\,, &\;\;\;\;\;\;\;\;\;\;&
\fr{\tilde a'}{\tilde a}(y_2) =  \fr{\kappa_5^2}{6} V_2\,,\\[2mm]\nonumber
\phi'(y_1)= \fr{1}{2} \fr{dV_1}{d\phi}\,,   &\;\;\;\;\;\;\;\;\;\;&
\phi'(y_2)= - \fr{1}{2} \fr{dV_2}{d\phi}\,.
\ea

 The final result takes the following form:
\be
\left. 2\tilde a_i^4 y_i\left [ {\cal L}_B(y_i) +\fr{\kappa_5^2}{6}V_i^2
-\fr{1}{4} \left( \fr{dV_i}{d\phi}\right)^2\right ] \right|_{i=1}^{i=2}
+
\left.\sum_{i=1,2} y_i \fr{\partial}{\partial(by_i)}U(by_1,by_2)\right|_{b=b_0=1}
=0.  
\label{E}
\ee
Here we used the smallness of the four-dimensional effective cosmological 
constant and put the first term in (\ref{first}) to zero. 
This algebraic constraint (as opposed to an integral condition (\ref{kkop-gkl}))
can be used to check the consistency of particular 
solutions. We note that in the particular cases that we study later, 
the separate cancellations occur for $y_1$ and $y_2$ proportional terms. 

It is easy to see that the {\it extremization constraint} is not uniquely defined.
Indeed, one can add to (\ref{E}) linear combinations of $T_5^5(y_i) 
- \kappa_5^2V_i^2/6 $ 
and ${\cal L}_B(y_i) - V_B(y_i) -(dV_i/d\phi)^2/8$,
 since these combinations are equal to zero 
due to the $55$-component of Einstein's equations and the boundary conditions
for  the scale factor and the scalar field:
\ba
\fr{1}{\kappa_5^2}\fr{\tilde a_i'^2}{\tilde a_i^2} = \fr{\kappa^2_5}{6}V_i^2 = 
T^5_5(y_i)\,,\nonumber
\\
{\cal L}_B(y_i) = \fr{1}{8}
\left(\fr {dV_i}{d\phi}\right )^2 + V_B(y_i)\,.
\label{55}
\ea

Using the above relations, we would like to give  another 
legitimate form of (\ref{E}):
\ba
\left. 2\tilde a_i^4 y_i\left [ V_B(y_i) + \fr{\kappa_5^2}{6}V_i^2
-\fr{1}{8} \left( \fr{dV_i}{d\phi}\right)^2\right ] \right|_{i=1}^{i=2}
+
\left.\sum_{i=1,2} y_i \fr{\partial}{\partial(by_i)}U(by_1,by_2)\right|_{b=b_0=1}
=0.  
\label{E1}
\ea

In order to obtain a radion mass term and impose a  
{\it stabilization constraint}, we differentiate 
$d\bar V_{eff}(b)/db$ over $b$:
\ba\label{second}
&&A^2\fr{d^2\bar V_{eff}}{db^2}=
\\\nonumber
&&
\left .
y_i^2\left[-\fr{12}{\kappa_5^2}(\tilde a'^2_i\tilde a^2_i)' + 
2 ({\cal L}_B(by_i)a_i^4)'\right ] \right|_{i=1}^{i=2}
+ \sum_{i=1,2} y_i^2 (  \tilde a^4_i V_i)'' 
+\sum_{i=1,2} y_iy_j\fr{\partial^2}{\partial(by_i)\partial(by_j)}U.
\ea
Not shown here are the terms proportional to the first and second derivatives 
of $A(b)$. These terms exist in principle, but vanish when $b=b_0$ because 
of $\Lambda_{eff}=0$ and because of the extremization constraint (\ref{E}),
and thus have no relevance for the calculation of the radion mass term. 
It is easy to see that the above expression contains $\tilde 
a''_i$. This may look 
worrisome as $\tilde a''$ has a singularity right at the position of the branes. 
However, in view of the definitions (\ref{defini}) the double derivatives 
in (\ref{second}) should be understood as the limit of the continuous part of 
$\tilde a''(y)$ at $y\rightarrow y_i$. 

As in the case of the {\it extremization constraint}, the expression 
(\ref{second}) can be significantly simplified at $b=b_0=1$, because of the 
use of the boundary conditions and the equations of motion for the scale 
factor and the scalar field. The extensive use of these relations allows 
to reduce (\ref{second}) to the following rather compact expression:
\ba
&&\left . A^2\fr{d^2\bar V_{eff}}{db^2} \right |_{b=b_0=1}=
\fr{4\kappa^2_5}{3}\sum_{i=1,2}y_i^2\tilde a_i^4 V_i\left[
T^5_5(y_i)+{\cal L}_B(y_i)- \fr{1}{2}\left(\fr{dV_i}{d\phi}\right)^2\right]
\label{S}\\\nonumber
&&
+\sum_{i=1,2}y_i^2\tilde a_i^4\left[ 
\fr{1}{4}\fr{d^2V_i}{d\phi^2}\left(\fr{dV_i}{d\phi}\right)^2
 - \fr{dV_i}{d\phi} \fr{dV_B}{d\phi}(y_i)\right]+
\left.\sum_{i=1,2} y_iy_j\fr{\partial^2}{\partial(by_i)\partial(by_j)}U
\right |_{b=1}\,.
\ea
The {\em stabilization constraint} is the requirement on the positivity of 
(\ref{S}). This result can be further modified to an equivalent form, 
using relations (\ref{55}) 
and the {\em extremization constraints} (\ref{E}) and (\ref{E1}). One of these 
forms contains only bulk and brane potentials and their derivatives over $\phi$:
\ba
&&\left . A^2\fr{d^2\bar V_{eff}}{db^2} \right |_{b=b_0=1}=
\fr{4\kappa^2_5}{3}\sum_{i=1,2}y_i^2\tilde a_i^4 V_i\left[
V_B(y_i) + \fr{\kappa^2_5}{6}V_i^2 -
\fr{3}{8}\left(\fr{dV_i}{d\phi}\right)^2\right]
\label{S1}\\\nonumber
&&
+\sum_{i=1,2}y_i^2\tilde a_i^4\left[ 
\fr{1}{4}\fr{d^2V_i}{d\phi^2}\left(\fr{dV_i}{d\phi}\right)^2
 - \fr{dV_i}{d\phi} \fr{dV_B}{d\phi}(y_i)\right]+
\left.\sum_{i=1,2} y_iy_j\fr{\partial^2}{\partial(by_i)\partial(by_j)}U
\right |_{b=1}\,.
\ea
The generalization of (\ref{E}), 
(\ref{E1}), (\ref{S}) and (\ref{S1}) to the case of multiple scalar fields
is straightforward.

\setcounter{equation}{0}

\section{Applications to brane-world models}
\medskip

{\em 1. No branes, trivial scalar field profiles.} \smallskip

In this case we take $V_i\equiv 0$, $\phi = const$, 
$V_B \equiv \Lambda_B$ and $\tilde a_i =1$. 
In the absense of branes, the individual positions $y_i$ lose their
meaning and translational 
invariance in $y$-direction requires that all quantities depend only on
the difference $y_2 - y_1$, which is simply related to the size of the
compact dimension. Therefore, 
$\bar V_{eff}(by_1,by_2) = \bar V_{eff}(b(y_2-y_1))$. Furthermore, since
$V_i = 0$, we also have $T_5^5 = 0$ (in the absence of branes this is
true for all $y$). Looking at the extremization constraints, we
immediately discover that 
\be
2 \Lambda_B\,(y_2-y_1) + \left. \fr{d U}{d b} \right |_{b=1}\, = 0.
\ee
The requirement $\Lambda_{eff}=0$ is equivalent to 
the condition $2\Lambda_B\,(y_2-y_1) + U =0$ that ensures the vanishing of 
the other components of $T_M^N$. The {\em stabilization constraint}
reduces to 
\be
\left . A^2\fr{d^2\bar V_{eff}}{db^2} \right |_{b=b_0=1}=
\left.\fr{d^2 U}{db^2}
\right |_{b=1}>0\,.
\ee
Particular examples of such a stabilization due to the Casimir force generated by 
multiple massive and massless scalar fields while classical bulk profiles 
of these fields are trivial were considered recently in Ref. \cite{EricPoppitz}.

\bigskip \medskip
\noindent
{\em 2. Randall-Sundrum model: empty bulk, empty branes, $U=0$.} \smallskip

In this case, the {\em extremization constraint} takes the form 
\be
\left. 2\tilde a_i^4 y_i\left [ \Lambda_B + 
\fr{\kappa_5^2}{6}\Lambda_i^2\right]\right |_{i=1}^{i=2}=0\,,
\label{RSE}
\ee
and leads to the usual relation between the bulk cosmological constant and 
the brane tension in the RS model: $-\Lambda_B=\kappa_5^2\Lambda_i^2/6$. 
The radion mass term, Eq. (\ref{S}) is also zero because 
$T_5^5 + {\cal L}_B = T_5^5 + \Lambda_B = 0$.
 It can actually be shown that every derivative of the radion effective
potential is trivially zero for this particular solution. Going back to
the expression of the potential, Eq. (\ref{eff1}), substituting
the relevant quantities inside the integral and performing the integration
over the extra coordinate, we are led to the result
\be
A^2\, \bar V_{eff}=
\left. -\sqrt{\frac{6}{\kappa^2_5 |\Lambda_B|}}\,\tilde a_i^4 \left [
\Lambda_B +
\frac{\kappa_5^2}{6}\Lambda_i^2\right]\right |_{i=1}^{i=2}
\ee
 We therefore conclude that every derivative of
the effective potential with respect to $b$, once evaluated for the specific
solution, will be identically zero. Therefore, this particular solution
is an {\it absolute saddle point} (a flat direction) of the radion
effective potential. 

\bigskip \medskip
\noindent
{\em 3. Empty branes, trivial scalar field profile, $U \ne 0$.}

\smallskip

Next, we turn to a particular two-brane-world solution in which 
the Casimir force generated by a conformally coupled scalar 
field was taken into account \cite{mukho,hkp}.
For the case of an attractive Casimir force, the $y$-profile of the 
scale factor takes the following form:
\be
\tilde a(t,y)=\cosh^{2/5}(\omega b y)\,,
\qquad \quad \omega^2=\frac{25}{24}\,\kappa^2_5 |\Lambda_B|\,.
\ee
For a repulsive Casimir force, the above cosh-like solution is substituted
by a sinh-like one and the analysis follows along the same lines.
The solution assumed a vanishing 4D cosmological constant, $\Lambda_{eff}=0$,
and no bulk scalar fields contributed classically to the five-dimensional
action. For both cases, the presence of the Casimir force is accounted for by 
the $U$ potential in the action:
\be
U(by_1,by_2) = - \fr{2\alpha}{L^4}\,,
\ee
where $\alpha$ is a dimensionless constant and $L$ the inter-brane distance
in 5D conformally-flat coordinates, defined as
\be 
L=\frac{1}{\omega}\,I(\omega b)= \frac{1}{\omega}\,
\int^{\omega by_2 }_{\omega by_1}\frac{d\xi}
{\cosh^{2/5}\xi}\,.
\ee

The {\em extremization constraint} (\ref{E1}) takes the following simple form:
\be
\left. 2\tilde a_i^4 y_i\left [ \Lambda_B + \fr{\kappa_5^2}{6}\Lambda_i^2 
+\fr{4\alpha \omega^5}{\tilde a_i^5 I^5(\omega) }\right ] \right|_{i=1}^{i=2}
=0. 
\label{ECasim}
\ee

Indeed, after applying the following relations from \cite{hkp}, 
\be 
\Lambda_i^2=\frac{6|\Lambda_B|}{\kappa_5^2}\,\tanh^2(\omega y_i)\,, \qquad
\quad |\Lambda_B| = \frac{4\alpha \omega^5}{I^5(\omega)}\,,
\ee
Eq. (\ref{ECasim}) is satisfied, separately for $y_1$- and
$y_2$-proportional terms. 

Using the above relations, we can calculate the radion mass term from 
Eq. (\ref{S1}):
\ba
\left. A^2\,\frac{d^2 \bar V_{eff}}{d b^2}
\right|_{b=1}=
-10 |\Lambda_B| \omega \Biggl\{\frac{1}{I(\omega )}
\biggl[\frac{y_2}{\tilde a_2} - \frac{y_1}{\tilde a_1}\biggr]^2
+ \frac{\kappa_5^2}{6} \biggl[\frac{y_2^2\Lambda_2}{\tilde a_2}+ 
\frac{y_1^2\Lambda_1}{\tilde a_1}\biggr]\Biggr\} < 0\,,
\ea
which coincides with the result of \cite{hkp} upon a trivial overall rescaling of 
$\tilde a$. The solution with the attractive Casimir force turns out to 
have a tachyonic radion, and therefore is unstable. 

Before concluding this subsection, we would like to briefly discuss
another solution with non-zero $U$ and with two positive tension
branes (with a negative bulk cosmological constant) which was derived in
\cite{kop1}. The conformal factor,
for this solution,  had the form 
\be
\tilde a(t,y)=\cosh^{1/2}(\omega b y)\,,
\qquad \qquad \omega^2=\frac{2}{3}\,\kappa^2_5 |\Lambda_B|\,.
\label{sol-T55}
\ee
The individual brane tensions (located at $y = y_1$ and $y=y_2$)
were given by
\be
\Lambda_i= \frac{3\omega}{\kappa^2_5} \tanh(\omega b y_i)\,,
\ee

In this case, the {\em extremization constraint} (\ref{E1}) becomes,
\be
\left. 2 \tilde a^4_i y_i (\Lambda_B  + {\kappa^2_5 \over 6} \Lambda_i )\,
\right|_{i=1}^{i=2} +
\left.\sum_{i=1,2} y_i \fr{\partial}{\partial(by_i)}U(by_1,by_2)\right|_{b=b_0=1}
=0,
\ee
and easily reduces to
\be
2 \Lambda_B (y_2 - y_1)  + 
\left.\sum_{i=1,2} y_i
\fr{\partial}{\partial(by_i)}U(by_1,by_2)\right|_{b=b_0=1} =0 .
\ee
As one can plainly see, the existence of this solution requires $dU/db
\ne 0$. In \cite{kop1}, this term was linked with $T_5^5$. 

Finally, we apply the {\em stabilization constraint} (\ref{S1}), which
becomes
\be
-2 \omega^2\,(y_1^2 \Lambda_1 + y_2^2 \Lambda_2)\, + 
\left.\sum_{i=1,2} y_iy_j\fr{\partial^2}{\partial(by_i)\partial(by_j)}U
\right |_{b=1}\,.
\ee
Thus provided that $d^2 U/db^2$ is positive and 
sufficiently large, this solution is stable. 

\bigskip \medskip
\noindent
{\em 4. Non-trivial $V_i(\phi)$, trivial bulk potential $V_B=\Lambda_B$, $U=0$.}
\smallskip

Exact solutions of this type, a massless scalar field in the bulk 
with non-trivial interactions on the branes  $V_i(\phi)$, were found in
Refs.  \cite{Stanford,kop1}. Originally, it was hoped that such a
solution had  some relevance for the cosmological constant problem, until
more careful  considerations \cite{Nilles} exemplified the necessity of 
fine-tuning in this model. For $y_1<y_2<0$, the solution was found
\cite{kop1} to be
 \be
\tilde a^4(by)=\sinh[\omega b\,(-y)]\,,
\qquad \qquad \fr{1}{b}\fr{d\phi}{dy} = \pm
\frac{\sqrt{2|\Lambda_B|}}{\sinh(\omega b\,|y|)}\,,
\label{sol-bulk}
\ee
where $\omega^2=8 \kappa^2_5 |\Lambda_B|/3$, and
the sign ambiguity is resolved only when the scalar potential
on either of the two branes is specified. 

The {\em extremization constraint} (\ref{E1}) takes the form
\be
\left. 2\tilde a_i^4 y_i\left [ \Lambda_B + \fr{\kappa_5^2}{6}V_i^2 
-\fr{1}{8}\left( \fr{dV_i}{d\phi}\right)^2\right ] \right|_{i=1}^{i=2}
=0,
\ee
and is trivially satisfied (separately for $y_1$ and $y_2$) 
upon the following substitutions:
\be 
V_i^2=\frac{6|\Lambda_B| }{\kappa_5^2}\,\coth^2(\omega |y_i|)\,,
\qquad
\fr{1}{4}\left(\frac{d V_i}{d \phi}\right )^2=  (\phi_i')^2 = \fr{2|\Lambda_B|}
{\sinh^2(\omega |y_i|)}.
\label{ft-bulk}
\ee
These relations allow one to express $dV_i/d\phi$ in terms of $V_i$ and
the  bulk cosmological constant, which obviously requires fine-tuning:
\be
\left( \fr{dV_i}{d\phi}\right)^2 = 8\left[ \Lambda_B + \fr{\kappa_5^2}{6} V_i^2
\right ].
\ee

Armed with these relations, we easily obtain the radion mass term from
(\ref{S1}), if we make one further assumption for
$d^2 V_i/d\phi^2$. In the case of  
linear dependence, $V_i = \alpha_i \phi_i$, 
the second line in (\ref{S1}) is zero and 
\be
\left. A^2\,\frac{d^2 \bar V_{eff}}{d b^2}
\right|_{b=1}= 
\fr{8\kappa^2_5}{3}\sum_{i=1,2}y_i^2\tilde a_i^4 V_i\left[
|\Lambda_B| - \fr{\kappa^2_5}{6}V_i^2\right]\,.
\label{radionmass}
\ee
By using the boundary conditions (\ref{ft-bulk}) twice, the radion mass
term may be written as
\be
\left. A^2\,\frac{d^2 \bar V_{eff}}{d b^2}
\right|_{b=1}= 
-4 \omega |\Lambda_B|\,\biggl[\,y_1^2\,\frac{\coth(\omega |y_1|)}
{\sinh(\omega |y_1|)} - y_2^2\,\frac{\coth(\omega |y_2|)}
{\sinh(\omega |y_2|)}\,\biggr]\,.
\label{radionmass-f}
\ee
The above expression is not sign definite. A simple numerical analysis
reveals that the sign of the radion mass squared depends on the inter-brane
distance (large inter-brane distances add a positive contribution
to the radion mass term), and the distance of the brane system from the
singularity located at $y_0=0$ (small separation between the branes
and the singularity adds a negative contribution). When 
$|y_1-y_0|=|y_1|<|y_1|_{crit}$, the stability of the system strongly
depends on the location of the second brane: if it is located closer to
the first brane rather than the singularity, the destabilizing
force is minimal and the solution is stable; if, however, the second
brane is moved closer to the singularity, the radion field is 
inevitably tachyonic. On the other hand, for $|y_1|>|y_1|_{crit}$, 
the stabilizing force coming from the large inter-brane distance
is dominant, which guarantees the overall stability of the system even
when the second brane is placed close to the singularity. 

For non-linear interaction terms $V_i$, there is an extra term in
the expression for the radion mass term proportional to
$d^2 V_i/d\phi^2$. If the sign of this term is chosen to be
positive then, for large inter-brane distances, this term tends
to stabilize the solution even more, while, for small brane separations,
it may reduce, or even overcome, the destabilizing force due to
the singularity.

As a final comment, let us stress at this point that the two-brane
solution with a bulk scalar field, studied above, turned out to be stable,
for much of the parameter space, in the absence of a 
bulk potential $V_B$ for the scalar field. This result differs from
that in \cite{GW}, where the mass of the radion is found to vanish in the 
limit of the vanishing bulk mass for the $\phi$ field. 
Our study shows that the non-trivial interaction terms
of the scalar field with the branes are sufficient for
both fixing the inter-brane distance \cite{kop1} and 
stabilizing the brane-system under small time-dependent
perturbations.

\sect{Conclusions}

In this paper, we have focused on the study of the stabilization of the 
size of extra dimensions. A variety of solutions in extra-dimensional
brane-world models have appeared in the literature that are characterized
by a constant radion field ($\dot b=0$). These solutions correspond
to extrema of the corresponding radion effective potential and
constraints involving the components of the energy-momentum tensor,
for the vanishing of the first derivative of $V_{eff}$, have been
written down for particular classes of models. 

In the context of our analysis, the derivation of those two types of 
constraints took place in the framework of the four-dimensio\-nal, 
effective theory for the graviton and radion field.
We demonstrated that the variation over the size of the extra 
dimension reduces the integral constraints to a set of algebraic
conditions  formulated at the positions of the branes. Clearly, the  
{\em extremization constraints} give the same information as 
Einstein's equations. However, the {\em stabilization constraint} is not
trivial. We were able to derive it in terms of the bulk and brane potentials,
which may be used for a direct calculation of the radion mass term in many 
brane-world models.  

The resulting constraints were
then applied for a number of static, five-dimensional solutions that
had previously appeared in the literature. All of them were shown to
satisfy the {\it extremization constraint} as expected, and furthermore
their stability behaviour was studied by making use of the {\it stabilization
constraint}. The two-brane Randall-Sundrum model was shown to be an
absolute saddle point (a flat direction) of the radion potential while
the implementation of an attractive Casimir force between the two branes
led to an unstable configuration, in agreement with a previous work.
Finally, the two-brane model with a bulk scalar field was shown to
be stable for large inter-brane distances, i.e. when the branes
lie in the exponential regime of the $\sinh$-like solution, while a
potential instability arises for small brane separations. Of particular
importance was the fact that the bulk potential is not a necessary ingredient 
for stabilizing an extra dimension with the scalar field. Indeed, 
the interaction terms between the bulk scalar field and the branes lead to
a  sufficient condition for stabilization.


{\bf Acknowledgements} 

P.K. and K.A.O. would like to thank Ian I. Kogan and
Antonios Papazoglou for helpful discussions at early stages of this work.
M.P. would like to acknowledge the support of PPARC. 
The work of  K.A.O.  was supported partly by DOE grant
DE--FG02--94ER--40823.

\end{document}